\documentclass[aps,prb,showpacs,twocolumn,superscriptaddress,floats]{revtex4}
\usepackage{amssymb}
\usepackage{amsbsy}
\usepackage{amsmath}
\usepackage{epsfig}

\newcommand{\srcubo}{SrCu$_2$(BO$_3$)$_2$ }

\begin{document}

\title {Variational treatment of the Shastry-Sutherland antiferromagnet using Projected Entangled Pair States (PEPS).}

\author{A. Isacsson}
\author{O. F. Sylju{\aa}sen}
\affiliation{NORDITA, Blegdamsvej 17, Copenhagen {\O}, DK-2100, Denmark}

\date{\today}

\begin{abstract}
We have applied a variational algorithm based on Projected Entangled Pair States (PEPS) to a two dimensional frustrated spin system, the spin-1/2 antiferromagnetic Heisenberg model on the Shastry-Sutherland lattice. We use the class of PEPS with internal tensor dimension $D=2$, the first step beyond product states ($D=1$ PEPS). We have found that the $D=2$ variational PEPS algorithm is able to capture the physics in both the valence-bond crystal and the Neel ordered state. Also the spin-textures giving rise to the magnetization plateaus seen in experiments on \srcubo are well reproduced. This shows that PEPS with the smallest nontrivial internal dimension, $D=2$, can provide valuable insights into frustrated spin-systems.
\end{abstract}

\pacs{}

\maketitle

\section{Introduction}
Spins with antiferromagnetic interactions prefer opposite alignment.  However, in many materials the lattice structure does 
not allow all antiferromagnetic bonds to be satisfied simultaneously. These are known as frustrated antiferromagnets and  
display a variety of different phases ranging from rather well-known Neel-ordered phases to much less understood 
exotic phases such as valence bond crystals and spin liquids. There is no general theory of frustrated antiferromagnets, 
thus the different lattice structures are usually studied as separate models. Especially difficult are 2D models. 
Finding the phase diagram of any of these models is made difficult by the lack of effective numerical tools that goes 
beyond exact diagonalization of the Hamiltonian. 

It is an unfortunate fact that the most powerful numerical methods such as the Density Matrix Renormalization Group (DMRG) 
and Quantum Monte Carlo (QMC) that are very effective for studying general quantum magnets do not work well when applied 
to frustrated 2D antiferromagnets. DMRG is mainly restricted to 1D systems, and QMC suffers from the sign-problem. 
However there are promising variational methods\cite{Nishino:2000,Nishino:2001,Maeshima:2001,Gendiar:2003,Maeshima:2004} 
that performs an energy minimization in a large class of states known as Tensor Product States (TPS). 
Recently Verstraete and Cirac suggested an alternative minimization strategy in this space of states, there 
termed Projected Entangled Pair States (PEPS), that promises to be very efficient\cite{PEPS_article} and deserves 
further study. The PEPS or TPS have a natural ``refinement'' parameter, the internal dimension $D$ of the tensors. 
This parameter determines how well the particular class of states covers the full Hilbert space. The lowest level $D=1$ 
corresponds to product states, thus yielding mean field theory results. The aim of the present article is to investigate 
how well the next level in the hierarchy, $D=2$, can describe a 2D frustrated antiferromagnet of real physical interest. 

\begin{figure}[t]
\epsfig{file=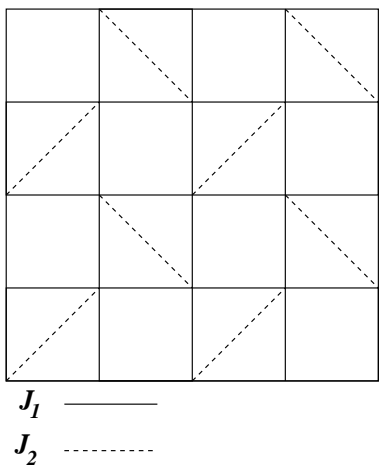,width=5cm}
\caption{Bond configuration for the Shastry-Sutherland model. All bonds have antiferromagnetic couplings. Vertical and 
horizontal bonds a coupling strength $J_1$ and diagonal bonds $J_2$.\label{Fig:SS_model}}
\end{figure}

An interesting frustrated antiferromagnet that exhibits both a Neel ordered phase and a valence bond crystal phase is the 
spin--1/2 Heisenberg antiferromagnet on the Shastry-Sutherland lattice, see Fig.~\ref{Fig:SS_model}. This model was 
initially proposed as a toy model possessing an exact dimerized eigenstate known as a valence bond crystal\cite{Shastry:1981}. 
However the interest in this model is more than academic as it is believed that the material compound SrCu$_2$(BO$_3$)$_2$ 
is reasonably well described by this model for particular values of the antiferromagnetic couplings\cite{Ueda_review:2003}. 
Although extensively studied, the zero temperature phase diagram of the Shastry-Sutherland antiferromagnet remains elusive. 
While two of the phases are known, the possible existence of an intermediate phase and its nature are still unresolved issues. 
In addition experiments on \srcubo in a magnetic field show the appearance of magnetization plateaus\cite{Kageyama:1999_1} 
with rather peculiar spin structures\cite{NMR}. Several theoretical approaches, based on the Shastry-Sutherland model have 
attempted to explain these 
steps~\cite{Miyahara:1999,Momoi:2000_1,Momoi:2000_2,Miyahara:2000,Miyahara:2000B,Fukumoto:2000, Fukumoto:2001, Girvin:2001, Miyahara:2003}. The approaches used so far have ranged from exact diagonalization~\cite{Miyahara:1999,NMR,Miyahara:2003}, 
perturbative analysis~\cite{Momoi:2000_1,Momoi:2000_2,Miyahara:2000,Miyahara:2000B,Fukumoto:2000, Fukumoto:2001} and 
mean field theory calculations~\cite{Girvin:2001}.

The PEPS or TPS are higher dimensional generalizations of Matrix Product States\cite{AKLT:1987,Fannes:1992} which are known to 
be particularly useful variational states in 1D\cite{Ostlund:1995}.  In contrast to the variational algorithm proposed in 
Ref.~\cite{PEPS_article} the variational calculations using TPS carried out in 
refs.~\onlinecite{Nishino:2000,Nishino:2001,Maeshima:2001,Gendiar:2003,Maeshima:2004} build in translational invariance at the 
outset in the minimization procedure by using site-independent tensors. While this reduces the number of variational parameters 
it is often desirable not to assume this when dealing with a spin system where the a priori unknown magnetic unit cell can 
be bigger than the unit cell of the lattice. There are also systems for which translational symmetry is explicitly broken 
by for instance impurities or boundaries. Thus it is desirable to have a method that is capable of treating also these 
situations. As a relevant example here, the NMR experiment on the 1/8 magnetization plateau in \srcubo  showed that the 
results were best explained in terms of a state that breaks translational symmetry\cite{NMR}.

From the viewpoint of 1D variational calculations where the MPS have internal matrix dimensions $D \sim 32-128$\cite{VMPS_impurities}, 
it would at first sight seem inadequate to restrict the 2D calculations to $D=2$. However as argued in 
Ref.~\onlinecite{PEPS_Computational_power} even the $D=2$ class of states is very rich, a fact that is supported 
by our findings. We find that the $D=2$ PEPS capture most of the known physics of the Shastry-Sutherland model such 
as the valence bond crystal phase, the Neel-ordered phase and the magnetization steps. 

The outline of this paper is as follows; In section~\ref{Sec:PEPS} we give a detailed outline of the variational method and 
in  Section~\ref{Sec:SS_model} we introduce the Shastry-Sutherland model and give a brief account of what is known about the 
ground state and the connection to the experimental results on \srcubo. In section~\ref{Sec:SS_PEPS_APP} we comment on the 
application of variational PEPS to the Shastry-Sutherland model and in sections~\ref{Sec:No_B_field} and~\ref{Sec:B_field} 
we look at the ground state with and without of external field respectively, examining the phase transition and 
magnetization plateaus. Finally in section~\ref{Sec:Numerical} we address the performance of the algorithm.

\section{Variational method using PEPS}
\label{Sec:PEPS}

\begin{figure}[t]
\epsfig{file=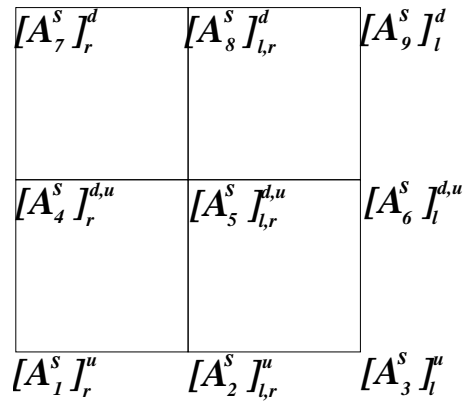,width=6cm}
\caption{Computational lattice and associated tensors. Shown here is an example of a $3\times 3$ computational lattice. 
With each site $i$ is associated 2 tensors $A_i^s$, $s=\uparrow,\downarrow$. Each tensor has indices $u,d,l,r$ 
corresponding to bonds connecting the site $i$ to neighboring sites. \label{Fig:comp_lattice}}
\end{figure}

Although the algorithm is described in Ref.~\onlinecite{PEPS_article} we reiterate it here in detail for completeness.  
As any variational algorithm, the aim is to minimize the expectation value of the Hamiltonian within a given class of trial states. 
The class of states used here are Projected Entangled Pair States
(PEPS) represented by an array of complex tensors $A_i$, each tensor associated with a physical spin. To define a PEPS trial wave function
an auxiliary lattice, the \emph{computational lattice}, is introduced. While the sites on the computational lattice coincide with the sites
on the physical lattice the bonds need not. However, it is important that the dimensionality of the computational lattice is the same
as the dimensionality of the physical lattice.  The bonds in the computational lattice determine the index structure of the tensors $A_i$.
A tensor $A_i$ will have one index for each bond in the computational lattice emanating out from site $i$. 
Note that different choices of the underlying computational lattice lead to different classes of variational PEPS-wave functions. 
As an example, consider Fig.~\ref{Fig:comp_lattice} where a computational lattice in the form of a simple $3\times 3 $ 
square lattice with open boundary conditions is shown. With each lattice site $i$ we then associate two tensors $A_i^s$, 
$s=\uparrow,\downarrow$ corresponding to spin up and spin down respectively. Each tensor has a rank determined by 
the number of bonds in the computational lattice connecting the site and a dimension $D$. Hence on site 5 in the lattice in  
Fig.~\ref{Fig:comp_lattice} we have two  ($s_5=\uparrow,\downarrow$) $D$-dimensional rank 4 tensors 
$[A_5^{s_5}]_{l,r}^{d,u}$ with indices for the bonds going down, up, left and right, whereas on site 6 
we have two rank 3 tensors $[A_6^{s_5}]_{l}^{d,u}$. While for PEPS one associates the tensor indices with the bonds in the computational lattice one could alternatively associate them with plaquettes as in the interaction-round-face TPS\cite{Sierra}.

For a system with $M$ sites we have the following form of the trial wave function
$$\left|\Psi\right>=\sum_{s_1=\uparrow,\downarrow}\cdots\sum_{s_{M}=\uparrow,\downarrow} {\mathcal Tr}\left(A_1^{s_1}\cdots A_{M}^{s_{M}}\right)\left|s_1\right>\cdots\left|s_{M}\right>$$
The symbol ${\mathcal Tr}(\cdot)$ means here that one should trace over all indices (bonds) in the computational lattice. As an example, for the $3\times 3 $ lattice in Fig.~\ref{Fig:comp_lattice} this operation becomes 
\begin{eqnarray}
{\mathcal Tr}(A_1^{s_1}A_2^{s_2}\cdots A_9^{s_9})&=& [A_1^{s_1}]_{r_1}^{u_1} [A_2^{s_2}]_{r_1,r_2}^{u_2}[A_3^{s_3}]_{r_2}^{u_3}\nonumber\\
&\times& [A_4^{s_4}]_{r_4}^{u_1 u_4} [A_5^{s_5}]_{r_4,r_5}^{u_2,u_5}[A_6^{s_6}]_{r_5}^{u_3,u_6}\nonumber\\
&\times& [A_7^{s_7}]_{r_7}^{u_4} [A_8^{s_8}]_{r_7,r_8}^{u_5}[A_9^{s_9}]_{r_8}^{u_6}
\end{eqnarray}
where repeated indices should be summed over. 

For $D=1$ the $A_i^s$:s are complex scalars and the trial wave function is a simple product state 
Ansatz similar to a mean field 
$$\left|\Psi_{D=1}\right>=\prod_{i=1}^M\sum_{s_i=\uparrow,\downarrow}A_i^{s_i}\left|s_i\right>.$$
For $D=2$ each index takes on two values. 
Although we will not make explicitly use of it in the following, each $A_i^{s_i}$ can for $D=2$ be represented as a vertex with arrows, one for each
index, each pointing either in or out. The contraction of all indices corresponds then to evaluating the partition 
function of a particular vertex model where $A_i^{s_i}$ represent the vertex weights\cite{Niggemann:1997}.

To minimize the energy (or to even calculate it) we need to evaluate 
$$\left<H\right>=\frac{\left<\Psi|H|\Psi\right>}{\left<\Psi|\Psi\right>}.$$
To see how this is done in practice we consider first the normalization $N=\left<\Psi|\Psi\right>$ with our $3\times 3$ 
example above which explicitly gives 
\begin{eqnarray}
\left<\Psi|\Psi\right>&=&\sum_{\{s'_i\}}\sum_{\{s_i\}}\left<\{s'\}\right| ([A_1^{s_1'}]_{r_1'}^{u_1'})^* ([A_2^{s_2'}]_{r_1',r_2'}^{u_2'})^*([A_3^{s_3}]_{r_2'}^{u_3'})^*\nonumber\\
&\times& ([A_4^{s_4'}]_{r_4'}^{u_1' u_4'})^* ([A_5^{s_5'}]_{r_4',r_5'}^{u_2',u_5'})^*([A_6^{s_6'}]_{r_5'}^{u_3',u_6'})^*\nonumber\\
&\times& ([A_7^{s_7'}]_{r_7'}^{u_4'})^* ([A_8^{s_8'}]_{r_7',r_8'}^{u_5'})^*([A_9^{s_9'}]_{r_8'}^{u_6'})^*\nonumber\\
&\times& [A_1^{s_1}]_{r_1}^{u_1} [A_2^{s_2}]_{r_1,r_2}^{u_2}[A_3^{s_3}]_{r_2}^{u_3}\nonumber\\
&\times& [A_4^{s_4}]_{r_4}^{u_1 u_4} [A_5^{s_5}]_{r_4,r_5}^{u_2,u_5}[A_6^{s_6}]_{r_5}^{u_3,u_6}\nonumber\\
&\times& [A_7^{s_7}]_{r_7}^{u_4} [A_8^{s_8}]_{r_7,r_8}^{u_5}[A_9^{s_9}]_{r_8}^{u_6}\left|\{s\}\right>
\end{eqnarray}
We now single out a specific site, say $k=5$, and construct the $D^2$ dimensional tensors $E_i, i\neq 5$
$$E_i=\sum_{s}(A_i^{s})^* \otimes (A_i^{s}).$$
Here the tensor product acts on all indices in the tensor, i.e., the tensors $E_j$ have composite indices
$$[E_j]^{\tilde{d},\tilde{u}}_{\tilde{l},\tilde{r}}=[E_j]^{(d'd),(u'u)}_{(l'l),(r'r)}.$$
The normalization can now be written as 
\begin{eqnarray}
\left<\Psi|\Psi\right>&=&\sum_{{s_5}}([A_5^{s_5}]_{r_4',r_5'}^{u_2',u_5'})^*[E_1]_{(r'_1r_1)}^{(u'_1u_1)} [E_2]_{(r'_1r_1),(r'_2r_2)}^{(u'_1u_2)}[E_3]_{(r'_2r_2)}^{(u'_3u_3)}\nonumber\\
&\times& [E_4]_{(r'_4r_4)}^{(u'_1u_1),(u'_4 u_4)}[E_6]_{(r'_5r_5)}^{(u'_3u_3),(u'_6u_6)}\nonumber\\
&\times& [E_7]_{(r'_7r_7)}^{(u'_4u_4)} [E_8]_{(r'_7r_7),(r'_8r_8)}^{(u'_5u_5)}[E_9]_{(r'_8r_8)}^{(u'_6u_6)} [A_5^{s_5}]_{r_4,r_5}^{u_2,u_5}\nonumber
\end{eqnarray}
Contracting all indices except those connecting site $k=5$ we get
\begin{eqnarray}
\left<\Psi|\Psi\right>&=&\sum_{{s_5},s'_5}([A_5^{s'_5}]_{r_4',r_5'}^{u_2',u_5'})^*\delta_{s'_5,s_5}[N_5]_{(r'_4r_4),(r_5'r_5)}^{(u'_2u_2),(u_5'u_5)}[A_5^{s_5}]_{r_4,r_5}^{u_2,u_5}\nonumber\\
\label{eq:someeq}
\end{eqnarray}
By treating the $2D^4$ components of $A_5$ as a $2D^4$ dimensional vector ${\bf A}_5$ and the $4D^8$ 
components of $\delta_{s'_5,s_5}N_5$ as a $(2D^4)\times(2D^4)$ matrix ${\mathcal N}_5^{\rm eff}$ Eq.~(\ref{eq:someeq}) becomes
$$\left<\Psi|\Psi\right>={\bf A}_5^\dagger {\mathcal N}_5^{\rm eff} {\bf A}_5.$$

The evaluation of $\left<\Psi|H|\Psi\right>$ can be done in a similar fashion if we treat each term in the Hamiltonian individually,
i.e, $H=\sum_n H^{(n)}$ where $H^{(n)}$ can be written as a product of on-site operators $H^{(n)}=\prod_{j=1}^M\hat{O}_j^{(n)}$.
Again we form $D^2$ dimensional tensors $E_j^{(n)}$ from the $D$ dimensional tensors $A_i^{s_i}$ by 
$$E_i^{(n)}=\sum_{s,s'}(A_i^{s'})^* \otimes (A_i^{s})\left<s'\right|\hat{O}_i^{(n)}\left|s\right>.$$
For each term $H^{(n)}$ we can now again write this in vector form if we single out a particular site $k$
$$\left<\Psi|H^{(n)}|\Psi\right>={\bf A}_k^\dagger {\mathcal H}^{(n)}_k {\bf A}_k$$
and sum the matrices ${\mathcal H}^{(n)}_k$ to obtain an effective Hamiltonian matrix 
${\mathcal H}_k^{\rm eff}=\sum_n {\mathcal H}^{(n)}_k$ for site $k$.

\begin{figure}[t]
\epsfig{file=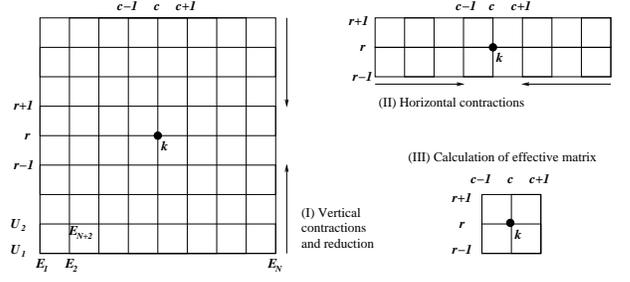, width=8cm, clip}
\caption{Steps in obtaining contribution to the effective operator matrices  ${\mathcal N}_k^{\rm eff}$ 
and ${\mathcal H}_k^{\rm eff}$ in Eq.~(\ref{eq:gen_ev}). For a site $k$ located at row $r$ and column $c$ we first 
contract row-wise from the top down to row $r+1$ and upwards from the bottom to row $r-1$ all the $E$-tensors contributing 
to the operator. After each row contraction the left-right dimension of the
tensors are reduced to a dimension $D_f$ before next row is contracted. When only rows $r$ and $r\pm 1$ remain we 
contract vertically to columns $c\pm1$ after which the effective matrices can be obtained.\label{fig:contr}}
\end{figure}

The overall structure of the optimization algorithm is now the following. We pick a site $k$ and
calculate ${\mathcal N}_k^{\rm eff}$ and ${\mathcal H}_k^{\rm eff}$ by contracting all indices of the $E$-tensors surrounding it.
Then we solve the generalized eigenvalue problem
\begin{equation}
{\mathcal H}^{\rm eff}_k {\bf A}_k=\lambda{\mathcal N}^{\rm eff}_k {\bf A}_k
\label{eq:gen_ev}
\end{equation} 
from which a new $A_k^s$ with lower energy can be determined. While it is in principle possible to chose this new $A_k^s$ to be the
eigenvector corresponding to the smallest eigenvalue in Eq.~(\ref{eq:gen_ev}) this occasionally leads to problems with convergence.
Instead we only gradually project out the high energy eigenvectors from ${\bf A}_k$ in the optimization. We then continue in this vein
sweeping over all sites $1\le k\le M$ until no further reduction in energy can be achieved. 

While it is no problem to obtain  ${\mathcal N}_k^{\rm eff}$ and ${\mathcal H}_k^{\rm eff}$ in the small $3\times 3$ example
above it becomes a problem as we move to larger systems. For a general set of tensors $E_i$ the trace 
${\mathcal Tr}(\prod_i E_i)$ is in its most general form, an NP-complete problem\cite{PEPS_Computational_power} 
and cannot be evaluated exactly for large systems but approximate strategies have to be used.
We have employed the strategy suggested in Ref.~\onlinecite{PEPS_article} doing this in a row-wise fashion.
To calculate either ${\mathcal N}^{\rm eff}_k$ or one of the contributions ${\mathcal H}^{(n)}_k$ the three steps in Fig.~(\ref{fig:contr})
are performed. For a site $k$ located at row $r$ and column $c$ we first contract vertically from the top down to 
row $r+1$ and upwards from the bottom to row $r-1$. When only rows $r$ and $r\pm 1$ remain we contract vertically to 
columns $c\pm1$ after which the effective matrices can be obtained.

When contracting vertically the left-right dimension of the $E$-tensors will increase. For instance contracting the $D^2$ dimensional 
tensors $[E_2]_{l,r}^u$ with $[E_{N+2}]_{l,r}^{d,u}$ in Fig.~(\ref{fig:contr}) generates a new tensor
\begin{equation}
[F_{2}]^u_{(l'l),(r'r)}=\sum_{x=1}^{D^2} [E_2]^{x}_{l',r'}[E_{N+2}]^{x u}_{l,r}
\label{eq:bleh}
\end{equation}
with $D^4$ dimensional left-right indices. This leads to an exponential growth of the left-right index dimensions with each 
row-contraction. To handle this we use the approximation technique suggested for approximating
an MPS with dimension $D_i$ with another MPS with a lower dimension $D_f\le D_i$ described in Ref.~\onlinecite{PEPS_article},
which has also been successfully used to simulate time-evolution in 1D systems\cite{DMRG_timeev}. Below we give
an example of how this is done for the contraction of row 1 with row 2 in Fig.~\ref{fig:contr}. 

\begin{figure}[t]
\epsfig{file=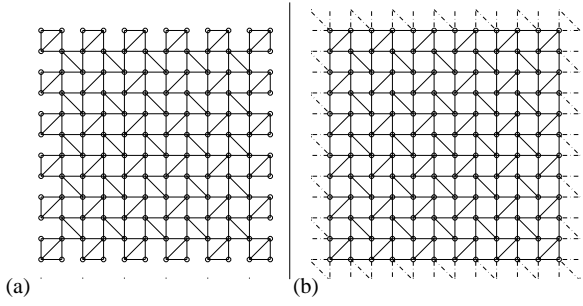, width=8cm, clip}
\caption{Boundary conditions used in the simulations. (a) Open boundary conditions. (b) Periodic BC. \label{fig:BC}}
\end{figure}

Any row in Fig.~\ref{fig:contr} can be viewed as a Matrix Product Operator (MPO). 
For the bottom row this corresponds formally to a vector
$$U_1=\sum_{\{u_i\}}E_1^{u_1}E_2^{u_2}\cdots E_N^{u_N}\left|u_1,\cdots,u_N\right>$$
represented by a set of $ND^2$ $D^2\times D^2$ matrices $E_i^u$.
One of the middle rows, for instance the second one, can be formally viewed as a matrix
$$U_2=\sum_{\{d_i\},\{u_i\}}E_{N+1}^{d_1,u_1}\cdots E_{N+N}^{d_N,u_N}\left|u_1,\cdots,u_N\right>\left<d_1,\cdots,d_N\right|.$$
Contracting row 1 with row 2 is thus formally equivalent to a vector-matrix multiplication giving rise to 
a new $ND^2$ dimensional vector $U_{21}=U_2U_1$ represented by $ND^2$ $D^4\times D^4$ matrices $F_i^u$ [cnf. Eq.~(\ref{eq:bleh})].
We now seek a new vector 
\begin{equation}
\tilde{U}_{21}=\sum_{\{u_i\}}\tilde{F}_1^{u_1}\cdots \tilde{F}_N^{u_N}\left|u_1,\cdots,u_N\right>
\label{eq:U21}
\end{equation}
represented by $ND^2$ $D_f\times D_f$ matrices ($D_f\le D^4$) $\tilde{F}_i^{u_i}$ such that
$$\kappa=|U_{21}-\tilde{U}_{21}|^2$$ is minimal. One does this in an iterative way starting with an Ansatz for the solution
and then optimizes the matrices $\tilde{F}_i^{u_i}$ one by one until convergence is reached.
In practice we do this by first forming the $D_f^2\times D_f^2$ matrices
$$G_i=\sum_u (\tilde{F}_i^u)^*\otimes \tilde{F}_i^u,$$ 
the $D_fD^4\times D_fD^4$ matrices
$$H_i=\sum_u (\tilde{F}_i^u)^*\otimes F_i^u$$ 
and the $D^8\times D^8$ matrices 
$$J_i=\sum_u (F_i^u)^*\otimes F_i^u$$ 
in terms of which $\kappa$ can be written
$$\kappa=\prod_i G_i-2{\rm Re}\prod_i H_i+\prod_i J_i.$$
For a given site $1\le k\le N$ along the row we can now get a linear equation for $\tilde{F}_k^u$ that
will locally minimize $\kappa$. To see this we differentiate with respect to $\left([\tilde{F}_k]^u_{l',r'}\right)^*$ 
\begin{eqnarray}
\frac{\partial\kappa}{\partial\left([\tilde{F}_k]^u_{l',r'}\right)^*} &=&\left[\prod_{i<k} G_i\right]_{l'l}[\tilde{F}_k]^u_{l,r}\left[\prod_{i>k}G_i\right]_{r'r}\nonumber\\
&-&\left[\prod_{i<k} H_i\right]_{l'l}[F_k]^u_{l,r}\left[\prod_{i>k}H_i\right]_{r'r}=0\nonumber,
\end{eqnarray} 
and treat the left-right indices of $F^u_k$ and $\tilde{F}^u_k$ as the indices of vectors ${\bf F}^u_k$ and ${\tilde{\bf F}^u_k}$  
which leads to the system of equations
$$G{\tilde{\bf F}}_k^u=H{\bf F}_k^u.$$
Although the matrices $J_i$ are not needed to actually do the minimization we still calculate them to keep control of the error.
If the error, after the $\tilde{F}_k^u$:s have converged is too large we increase $D_f$ to obtain a better approximation.

For an $N\times N$ system we need to calculate of the order of $N^2$ contributions ${\mathcal H}^{(n)}$ to the effective Hamiltonian
${\mathcal H}^{\rm eff}$. For each contribution the contractions and approximations of MPO:s [cnf. Fig~\ref{fig:contr}] 
need to be calculated. This is the most computationally costly part of the algorithm and to avoid 
unnecessary calculations we optimize the tensors
$A_i^s$ in the computational lattice row-wise and store all calculated MPO:s which can be reused. This means that 
the memory needed for storage of MPO:s scales as $N^4 D^2 D_f^2$.

Finally we would like to point out that in this implementation we make no use whatsoever of any symmetries of the Hamiltonian,
neither in algorithm nor in the trial states. This means that our program can treat very general Hamiltonians with 
nonuniform ground states.

\section{Shastry-Sutherland Model}
\label{Sec:SS_model}
The Shastry-Sutherland model was originally introduced as an example of a model with an exact dimerized ground state\cite{Shastry:1981}.
The model is a frustrated spin-1/2 antiferromagnet with a bond configuration shown in Fig.~\ref{Fig:SS_model}
\begin{equation}
H=J_1\sum_{\left<i,j\right>}{\bf S}_i\cdot{\bf S}_j+J_2\sum_{\left<i,j\right>'}{\bf S}_i\cdot{\bf S}_j.
\label{Eq:SS_ham}
\end{equation}
Although the model was introduced for reasons of purely theoretical nature, interest was renewed along with 
experiments on SrCu$_2$(BO$_3$)$_2$~\cite{Kageyama:1999_1}. In SrCu$_2$(BO$_3$)$_2$ the crystal structure is layered with 
alternating planes of CuBO$_3$ and Sr and the magnetic properties stem from the CuBO$_3$-layers. It has been 
argued that these layers are well modeled by the Shastry-Sutherland model~\cite{Ueda_review:2003}.

In the limit $J_2\gg J_1$ the ground state, which is separated from the excited states by a gap, is a dimer state 
with localized spin singlets on the diagonal bonds, the ground state energy per spin being $E_{\rm dimer}=-3/8 J_2$ per 
diagonal bond. In the other limit $J_1\gg J_2$ the model reverts to the ordinary antiferromagnetic 
Heisenberg model with a Neel ordered ground state and gapless spectrum. From high temperature series 
expansion and exact diagonalization~\cite{Miyahara:1999,Weihong:1999,Muller-Hartmann:2000} a possible direct transition 
between the dimer phase to the Neel phase has been estimated to lie at $(J_1/J_2)_c=0.7\pm 0.01$.

\begin{figure}[t]
\epsfig{file=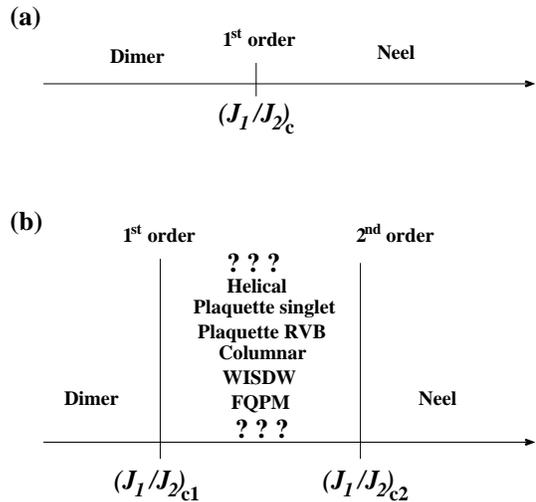,width=7cm}
\caption{Proposed phase diagrams for the Shastry-Sutherland model. Two possible scenarios for the phase diagram
have been proposed. Either a direct transition between the dimer state and the Neel state (a) or a transition via an intermediate
phase as in (b). The nature of this intermediate phase has not been established.\label{Fig:SS_PD}}
\end{figure}

Other works point to the existence of an intermediate phase between the antiferromagnet and the dimer phase.
A sketch of the phase diagram is shown in Fig.~\ref{Fig:SS_PD}. Most estimates agree that below $(J_1/J_2)_{c1}>0.6$ the 
ground state is the dimer state and above $(J_1/J_2)_{c2}<0.9$ the ground state is the Neel state.
The nature of this intermediate state has been addressed in several publications. In Ref.~\onlinecite{Albrecht:1996} Albrecht and 
Mila used Schwinger Boson mean field theory to argue in favor of a first order transition between the dimer state into a 
helical state and a second order transition to a Neel state. Another possible intermediate state, the plaquette singlet 
phase was discussed by Koga and Kawakimi in Ref.~\onlinecite{Koga:2000}. Both plaquette states and Helical 
states were considered in Ref.~\onlinecite{Sachdev:2001}. Arguments against both plaquette and helical phases was 
put forward in Ref.~\onlinecite{Weihong:2002}
who performed extensive series expansions around both the helical and plaquette phases and even columnar phases. This is 
in contrast to Ref.~\onlinecite{Hajj:2005} which supports either a plaquette phase or a columnar phase. Other suggestions 
for the intermediate phase are Weakly Incommensurate Spin-Density Waves (WISDW) or 
Fractionalized Quantum Para-Magnet (FQPM)~\cite{Carpentier:2001}. Finally a resonant valence bond plaquette 
phase was suggested as the intermediate state in Ref.~\onlinecite{Wessel:2002}. Thus, neither the existence 
of an intermediate phase nor its exact nature are presently known.

Experiments on SrCu$_2$(BO$_3$)$_2$ in strong external fields show the existence of magnetization 
plateaus~\cite{Kageyama:1999_1,Ueda_review:2003}. While
the ground state in absence of an external field is believed to be the dimer-state, the magnetization steps were originally thought 
to be formed by strongly localized triplets forming a periodic patterns which may spontaneously break the 
translational symmetry~\cite{Momoi:2000_1,Momoi:2000_2,Miyahara:2000,Miyahara:2000B,Fukumoto:2000, Fukumoto:2001} 
(for an alternative explanation hypothesis see Ref.~\onlinecite{Girvin:2001}). Subsequent NMR experiments~\cite{NMR} 
at the 1/8 plateau revealed
a more complex structure, inconsistent with the simple triplet-singlet picture. By including coupling to phonon degrees, with the sole
purpose of breaking the translational symmetry, exact diagonalization studies of small systems~\cite{NMR,Miyahara:2003} 
revealed more complex spin textures. Again, at all steps (except the 1/2) translational symmetry is broken and 
larger unit cells are formed.

\section{Application of variational PEPS to the Shastry-Sutherland model}
\label{Sec:SS_PEPS_APP}
Applying variational PEPS to the Shastry-Sutherland model is straight forward. 
Two issues should be noted. First, we stress again that the computational lattice does not need to have the 
same bond-configuration as the underlying Hamiltonian.
For the purpose of studying the Shastry-Sutherland model it is sufficient to use an ordinary square lattice 
as depicted in Fig.~\ref{Fig:comp_lattice}. It is easy to show that already with a low tensor dimension $D=2$ it is 
possible to represent exactly the dimerized ground state with singlets on all diagonal bonds. 

The second issue regards boundary conditions. In implementing the algorithm we have used a computational lattice with 
open boundary conditions. The reason for this is two-fold. Firstly, using a computational lattice with periodic 
boundary conditions severely reduces the performance of the algorithm, the difference being that between matrix-vector 
multiplications rather than matrix-matrix multiplications. Secondly, 
our program suffers from stability problems arising due to ill-conditioning and round-off errors in the case of computational lattices
with periodic boundary conditions.

Although, the computational lattice does not have periodic boundary conditions it is not necessary to adopt the same 
boundary conditions to the Hamiltonian. In this study we have used two different physical boundary conditions, open and periodic.
For the open BC we have adopted the geometry shown in Fig.~\ref{fig:BC}(a) while for the periodic the geometry in Fig.~\ref{fig:BC}(b).
The interpretation of using periodic BC in the physical problem while using open BC in the computational lattice is reminiscent of
using a self-consistent field on the boundary. The bonds across the boundary have only a tensor dimension $D_{\rm boundary}=1$ 
which in the limit of large lattices implies that the contribution from a such a bond will approach the product form 
$\left<{\bf S}_i\cdot{\bf S}_j\right>\rightarrow\left<{\bf S}_i\right>\cdot\left<{\bf S}\right>_j$
for limited $D$. 

We have mainly restricted ourselves to using a tensor-dimension $D=2$ for which a usual work-station with 1GB of internal 
memory suffices. Although our program can in principle handle $D>2$ (see section.~\ref{Sec:Numerical}), it is in its 
present incarnation too slow and unstable for $D>2$. For the dimension of effective MPO:s when calculating effective 
operators we have used a variable $16\le D_f\le 24$ for all simulations except for the largest ($12\times 12$) 
systems where the 1GB memory limit restricts us to $16\le D_f\le 18$.  

\section{Ground state in zero field.}
\label{Sec:No_B_field}

\begin{figure}[t]
\epsfig{file=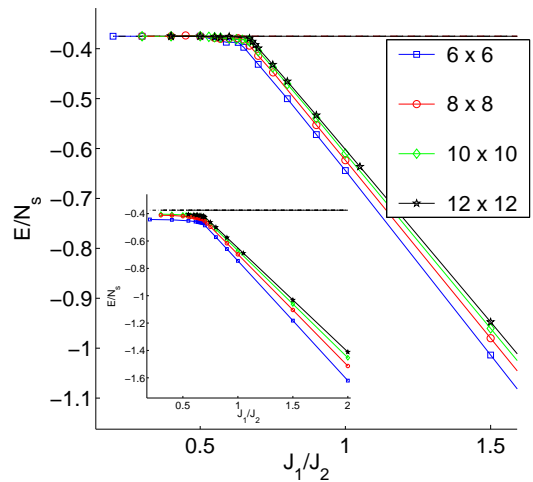, width=7cm}
\caption{(Color online) Variational minimum energy as a function of $J_1/J_2$ for system sizes  
$6\times 6$ (blue squares), $8\times 8$ (red circles), $10\times 10$ (green diamonds) and $12\times 12$ (black stars). 
For open boundary conditions (main figure) the transition from
the dimer state to the Neel state is clearly visible in the energy which has been scaled to the number of diagonal bonds. The inset shows
energies for periodic boundary conditions using the same energy scaling.\label{fig:EN_OBC}}
\end{figure}

In Fig.~\ref{fig:EN_OBC} the lowest energies obtained by the algorithm for $D=2$ are shown for system sizes of
$6\times 6$, $8\times 8$, $10\times 10$ and $12\times 12$ with both open (main panel) and periodic boundary conditions (inset).
The total energy has been scaled by the number of internal
diagonal bonds $N_s$ and the coupling energy $J_2$. As can be seen, for low $J_1/J_2$ the ground state energy approaches -3/8$J_2$ for the system 
with open boundary conditions confirming the convergence to 
the dimerized ground state. From the graph obtained using open boundary conditions it can be seen that 
already for $D=2$ we find a phase transition from the dimerized state. The transition point being located at $0.69\pm0.02$ 
which is in agreement
with estimates for the transition point of the direct dimer-Neel transition.
Further, in the energy a clear finite size effect is seen as the transition point is approached.

\begin{figure}[t]
\begin{center}
\epsfig{file=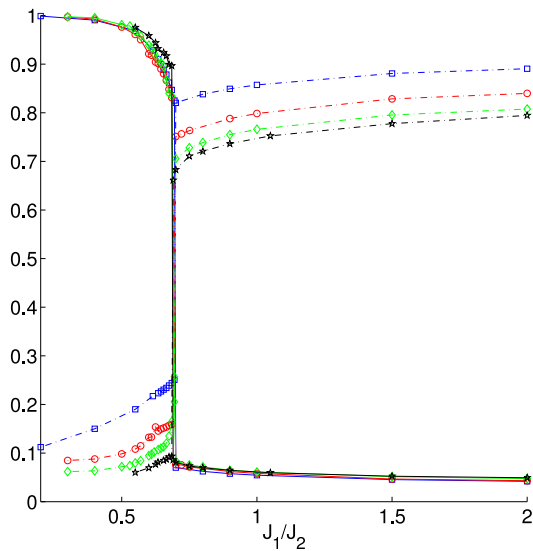, width=8cm,clip}
\vspace*{-2.6cm}
\end{center}
\caption{(Color online) Order parameters, singlet mixing on diagonal bonds (solid lines) and staggered magnetization 
(dashed lines) for periodic boundary conditions for system sizes 6x6 (blue squares), 8x8 (red circles), 10x10 (green diamonds) 
and 12x12 (black stars). Results obtained with internal tensor dimension $D=2$ suggests a direct first order transition 
from the dimer-state to the Neel state.
\label{fig:EN_PBC}}
\end{figure}

From the inset showing the energies obtained using periodic boundary conditions the location and nature of the transition is less clear. 
Here we have again divided the total energy by the number of 
internal diagonal bonds. Since the diagonal bonds across the boundary are not counted this gives rise to energies lower than $-3/8J_2$.
The transition is revealed by looking at the strength of the diagonal singlets and the staggered magnetization, 
as shown in Fig.~\ref{fig:EN_PBC}. The singlet mixing is calculated by projecting the diagonal bonds on to the singlet state, i.e.
1 indicates a singlet while 0 indicates a triplet. The staggered magnetization displayed is calculated as 
$2{<(\sum_i {\bf S}_i (-1)^i)^2>^{1/2}}$. Note that for large $J_1/J_2$ the staggered magnetization is higher than the value expected 
for a Heisenberg antiferromagnet. The reason for this can be two-fold. Firstly, finite size effects explain a 
part of the discrepancy as can be seen from the graph. Secondly, while $D=2$ gives a good value for the energies involved, 
being only a few percent off the exact values, observables may differ by more (see Sec.~\ref{Sec:Numerical}).

As stated in Section~\ref{Sec:SS_model}, there are good reasons to believe that an intermediate phase exists between the dimer phase
and the Neel phase. However, it is hardly surprising that we don't see this intermediate phase in our $D=2$ variational calculation.
First of all, with tensor dimension $D=2$ we typically overshoot the true ground state energy by a few percent, thus higher
$D$ is likely needed to capture any additional phase that may differ by a percent or less in energy. Second, the influence of boundary conditions scales as $1/N$ which implies that boundary effects 
can have a big impact even for the largest systems ($12\times12$) in cases where the energy splittings 
between ground state candidates are small.

\section{Magnetization Plateaus}
\label{Sec:B_field}

\begin{figure}[t]
\epsfig{file=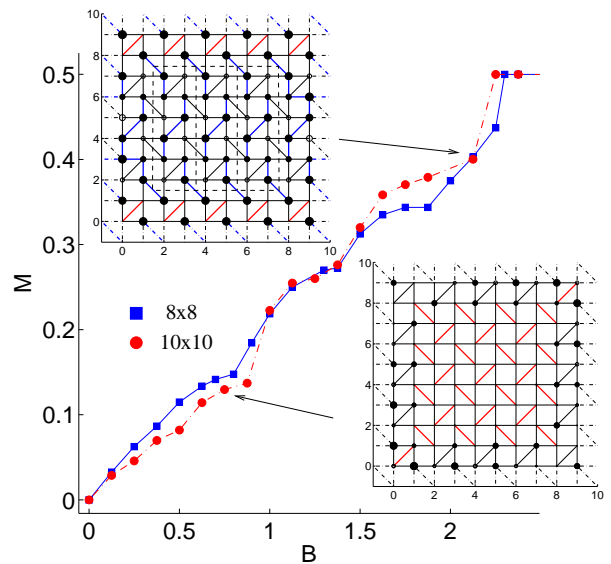, width=8cm}
\caption{(Color online) Magnetization curve at $J_1/J_2=0.6$ for system sizes $8\times 8$ (blue squares) and 
$10 \times 10$ (red circles). A clear step like structure corresponding to fillings 1/4,1/3 and 1/2 are seen. The insets 
show the distribution of singlets (red lines), triplets (blue lines). The $S_z$ component of the spin on each site 
illustrated by a circle with radius proportional to $|\left<S_z\right>|$. Filled circles are aligned with the field 
while open circles represent spins pointing opposite to the field.\label{fig:magncurve}}
\end{figure}

To study the magnetization curve we have deliberately chosen a somewhat smaller coupling 
constant $J_1/J_2=0.6$ than the experimental value $(J_1/J_2)_{SrCu_2(BO_3)_2}=0.635$. This makes the dimer state more stable and the 
algorithm converges faster but should not significantly affect the physics.
In Fig.~\ref{fig:magncurve} we show the results for a finite magnetic field for systems with periodic boundary conditions 
of sizes $8\times 8$ and $10\times 10$ ($D=2$). Although our square geometry and (periodic) boundary 
conditions are inconsistent with the unit cells proposed in Refs.~\onlinecite{NMR,Miyahara:2003} a clear step like structure is nevertheless visible in Fig.~\ref{fig:magncurve}. 
A closer inspection reveals that \emph{locally} the spin configurations we have obtained 
match those in Ref.~\onlinecite{Miyahara:2003} very well.

To study the spin configurations at the plateaus we have visualized the wave functions by coloring the bonds according 
to the amount of triplet or singlet mixing (see Fig.~\ref{fig:magncurve}). Localized singlets are drawn in red 
whereas triplets are blue. To determine the color of a bond we use the following criterion
\begin{equation}
\left\{
\begin{array}{l|c}
{\rm red} & \left<\Psi\right|{\bf S}_i\cdot {\bf S}_j \left|\Psi\right> <0.1 \epsilon_{\rm triplet}+0.9 \epsilon_{\rm singlet} \\
{\rm blue} & \left<\Psi\right|{\bf S}_i\cdot {\bf S}_j \left|\Psi\right> >0.9 \epsilon_{\rm triplet}+0.1 \epsilon_{\rm singlet} \\
{\rm black} & {\rm otherwise}
\end{array}
\right.
\end{equation}
Furthermore we have measured the spin component parallel to the direction of the applied field (${\bf B}=B\hat{\bf z}$) and visualized
$\left<S_z\right>$ by circles with radii proportional to $|\left<S_z\right>|$. Spins aligned (anti-aligned) with 
the field are drawn as filled (open) circles.

For small fields, $B<1$, we see a finite magnetization where one would expect a spin gap. This is due to 
the inability of our trial states to form singlets across the boundary. As can be seen in the inset, for $B<1$ the interior 
of the system is still in the dimer phase while only spins on the boundary have aligned with the field. Thus, by 
looking at when the magnetization in the interior of the system becomes finite we estimate the spin gap to be 
roughly $B=1$ corresponding to 33 T where we have used coupling constants $J=85$ K, $g=2.28$ (with J=71K the corresponding 
number is 28 T). 

For $B>1$ three steps can be distinguished, 1/4, 1/3 and 1/2. While the spin texture at 1/2 matches that of earlier predictions, half
of the diagonal bonds being triplets while the other half being singlets, the spin textures for other points are 
more elaborate and only agrees with earlier predictions locally. An example is
shown in the top left inset of Fig.~\ref{fig:magncurve}. Here, three of the 1/3 unit cells obtained 
by Miyahara et al~\cite{Miyahara:2003} are reproduced in the interior of the $10\times 10$ system. Note that the total magnetization is larger than 1/3
at this point due to the spin configurations on the boundary.

The spin textures obtained at the steps can only be found when translational symmetry is broken. In the variational method employed here translational symmetry is broken partly because of our choice of computational 
lattice and boundary conditions, and partly because a $D=2$ PEPS cannot represent the coherent superposition of degenerate plateau states connected by global symmetry transformations. 

\section{Algorithm performance}
\label{Sec:Numerical}
\begin{figure}
\epsfig{file=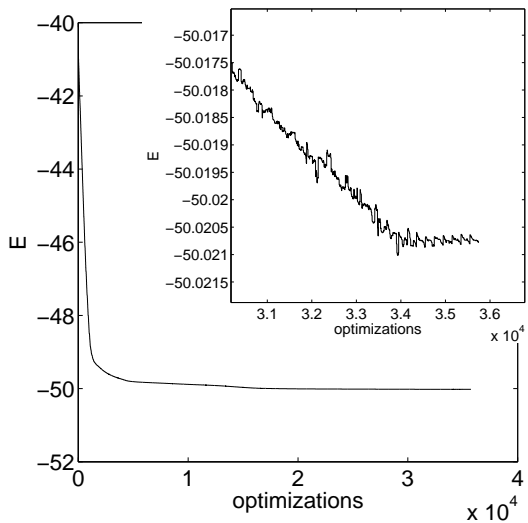, width=7cm}
\caption{Convergence of energy in a $10\times 10$ system with periodic boundary conditions for $D=2$. The energy is shown
as a function of the number of diagonalizations of the generalized eigenvalue problem in Eq.~\ref{eq:gen_ev}. The inset shows
a closeup of the final part of the optimization where fluctuations due to the approximation strategy used to calculate expectation
values are visible.\label{fig:convergence}}
\end{figure}

\begin{figure}
\epsfig{file=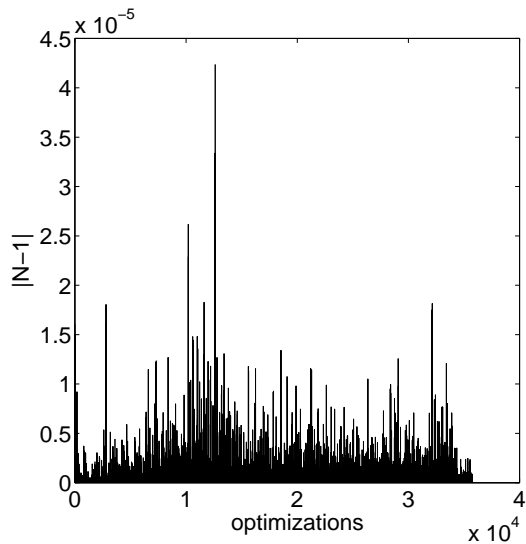,width=7cm,clip}
\caption{Deviation of the norm from unity during the optimization of a $10\times 10$ system with $J_1=J_2$.\label{fig:normdev}}
\end{figure}

So far we have only concerned ourselves with $D=2$ which is the first step beyond a simple on-site factorizable wave function.
Restricting ourselves to $D=2$ and sizes up to $12\times 12$ allows the program to run on an ordinary workstation with 1GB internal memory
without using any swapping to disk.

Because the method is based on a sequence of approximations, i.e., for a 12x12 system there are over 700 contributions to 
the effective Hamiltonian on any given site, each contribution being obtained in a series of up to 10 consecutive approximations
one has to ask whether or not the precision is compromised. Another important factor to consider is how much more accuracy 
(how much closer to the true ground state energy we can come) can be obtained by increasing $D$, and how the 
computational effort scales with increasing $D$.

In Fig.~\ref{fig:convergence} the energy as a function of number of optimizations is shown for a $10\times 10$ system with $J_1=J_2=1$.
While smooth on a large scale, the errors accumulated in the successive approximations are clearly visible in the inset which shows
a closeup of the final convergence. In this simulation and others we have used a final dimension in the
approximation of $E$-tensors (see Sec.~\ref{Sec:PEPS}) $16\le D_f\le 24$ (For $12\times 12$ we have been 
restricted to $16\le D_f\le 18$  due to the limited memory (1GB) of the workstation). As can be seen, despite the 
heavy reduction of the state space in the calculations, we have a precision of the order of 4 digits. This can also be seen by 
looking at the norm of the wave function. In Fig.~\ref{fig:normdev} the deviation of the norm from the 
nominal value 1 is shown for the simulation in Fig.~\ref{fig:convergence}.

\begin{figure}[t]
\epsfig{file=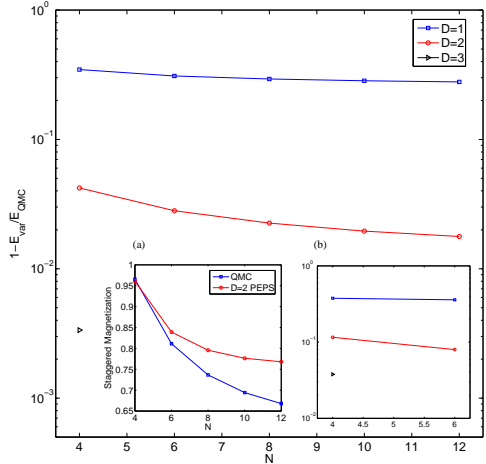,width=10cm,clip}
\caption{(Color online) Comparison between ground state energies obtained by variational PEPS and Quantum Monte Carlo 
for open Heisenberg antiferromagnet with open boundary conditions for system sizes $4\times 4$ to $12\times 12$ and 
tensor dimensions ranging from $D=1$ to $D=3$. Inset (a) shows a comparison 
between QMC and $D=2$ PEPS for the staggered magnetization for the Heisenberg model. Inset (b) shows a comparison between exact diagonalization of the 
Shastry-Sutherland model with periodic boundary conditions and PEPS $D=1$ to $D=3$.  \label{fig:QMC_comp}}
\end{figure}

To estimate the accuracy, i.e. how close to the true ground state energy the variational PEPS algorithm can get we 
have compared it to QMC on an ordinary Heisenberg antiferromagnet (QMC being unable to handle the SS-model), 
$H=J\sum_{\left<i,j\right>}{\bf S}_i\cdot{\bf S}_j$
with open boundary conditions. The QMC was run at an inverse temperature of $\beta=256/J$ and the results
are shown in Fig.~\ref{fig:QMC_comp} where the quantity $1-E_{\rm PEPS}/E_{\rm QMC}$ is shown for different system sizes and 
tensor dimensions $D$. For the smallest system $4\times 4$ sites we find good agreement with the figures reported in 
Ref.~\onlinecite{PEPS_article} obtained using imaginary time evolution. We further note that as the system size is 
increased the relative error decreases slightly, and that a linear increase in $D$ seems to give an 
exponential increase in accuracy. 

We want to point out that although the energies obtained by PEPS are in good agreement  with exact results, observables 
may deviate more. In the left inset of Fig.~\ref{fig:QMC_comp} the staggered magnetization for the Heisenberg model 
is compared with the staggered magnetization obtained using $D=2$ PEPS. Compared to the accuracy in energy which 
is of the order one percent (see main panel) the error in the staggered magnetization is an order of magnitude larger.

Although the above comparison for the Heisenberg model does not, in a strict sense, tell us 
anything about the accuracy obtained for the SS-model away from the limit $J_2/J_1\gg 1$ it still serves as a good indication
on the general behavior as one varies $N$ and $D$. For the SS-model we have compared with exact diagonalization results obtained 
using SPINPACK~\cite{spinpack,Richter:2004} at $J_1=J_2$ using periodic boundary conditions. The comparison is shown 
in the right inset of Fig.~\ref{fig:QMC_comp}.

The internal tensor dimension $D$ plays an important role in how faithfully a PEPS can represent ground states. The 
algorithm scales very badly with increasing $D$, the bottle neck being the contraction of two rows 
(See Eq.~\ref{eq:U21}). To form an MPO by contracting two rows requires of the order $ND^6 D_f^2$ ($D_f>D^2$) and 
scaling with $D$ is at best $D^{10}$. This then sets a limit to the maximum internal dimensions that can be practically 
used and the large values of $D\sim 10^2$ used in 1D variational MPS cannot be reached. However, as we have 
seen here (see also Ref.~\onlinecite{PEPS_Computational_power}), for 2D PEPS we expect small $D\sim 2-5$ to be able 
capture the essential physics for many problems with short range interactions. The scaling of the algorithm with linear system size is $N^4$ and is less severe.

All simulations were run on Linux workstations with 2.0 GHz 
AMD Athlon processor and 1GB of internal memory. On such a machine the graph in Fig.~\ref{fig:convergence} took 96h to produce.

Finally we comment on the stability of the algorithm. We have found that the technique used to optimize tensors one by one
often becomes unstable and may not be the optimal way to find the ground state. It may well be that using imaginary time
evolution is a more effective way. 

\section{Conclusions}
We have applied a variational procedure based on Projected Entangled Pair States (PEPS) to study the ground 
state properties of a frustrated spin system, the
Shastry-Sutherland model. Using the smallest nontrivial dimension on the tensors $D=2$ a direct phase transition 
between the dimer state and the Neel state can be observed, the location being well in agreement with other theoretical estimates
for a direct transition. Within $D=2$ we see no clear indication of an intermediate phase which may require 
higher $D$, larger systems, or proper handling of periodic boundary conditions.
We also find that already with PEPS $D=2$, magnetization plateaus are possible to reproduce, and that the non-trivial spin textures associated with these plateaus can be seen.  

Furthermore we have examined the performance of the algorithm and conclude that  it degrades rather severely for intermediate to large values of the internal dimension $D$. However, this scaling of the performance degradation might not be so restrictive as already the $D=2$ class of PEPS is well suited for studies of frustrated spin systems at a level beyond mean field theory. The method can readily be extended to other 2D frustrated spin-models.

\end{document}